\documentclass[conference]{IEEEtran}
\IEEEoverridecommandlockouts
\usepackage[utf8]{inputenc}
\usepackage{cite}
\usepackage{amsmath,amssymb,amsfonts}
\usepackage{algorithmic}
\usepackage{graphicx}
\usepackage{textcomp}
\usepackage{xcolor}
\usepackage{color}
\usepackage{booktabs}

\makeatletter
\@ifundefined{showcaptionsetup}{}{
 \PassOptionsToPackage{caption=false}{subfig}}
\usepackage{subfig}
\makeatother

\usepackage{eso-pic}
\newcommand\AtPageUpperMyright[1]{\AtPageUpperLeft{
 \put(\LenToUnit{0.5\paperwidth},\LenToUnit{-1cm}){
     \parbox{0.5\textwidth}{\raggedleft\fontsize{9}{11}\selectfont #1}}
 }}
\newcommand{\conf}[1]{
\AddToShipoutPictureBG*{
\AtPageUpperMyright{#1}
}
}

\usepackage{tikz}
\usepackage{textcomp}
\usepackage{hyperref}
\newcommand\copyrighttext{%
  \footnotesize \textcopyright 2021 IEEE.  Personal use of this material is permitted.  Permission from IEEE must be obtained for all other uses, in any current or future media, including reprinting/republishing this material for advertising or promotional purposes, creating new collective works, for resale or redistribution to servers or lists, or reuse of any copyrighted component of this work in other works.
  DOI: Pending Release by IEEE}
\newcommand\copyrightnotice{%
\begin{tikzpicture}[remember picture,overlay]
\node[anchor=south,yshift=10pt] at (current page.south) {\fbox{\parbox{\dimexpr\textwidth-\fboxsep-\fboxrule\relax}{\copyrighttext}}};
\end{tikzpicture}%
}

\def\BibTeX{{\rm B\kern-.05em{\sc i\kern-.025em b}\kern-.08em
    T\kern-.1667em\lower.7ex\hbox{E}\kern-.125emX}}

\title{Delayed Asynchronous Iterative Graph Algorithms}

\author{
    \IEEEauthorblockN{Mark P. Blanco\IEEEauthorrefmark{1}, Scott McMillan\IEEEauthorrefmark{2}, Tze Meng Low\IEEEauthorrefmark{1} \\
    \textit{\IEEEauthorrefmark{1}Dept. of Electrical and Computer Engineering ~~~~~ \IEEEauthorrefmark{2}Software Engineering Institute} \\
    \textit{Carnegie Mellon University}\\
    Pittsburgh, PA, United States \\
    \{markb1, scottmc, lowt\}@cmu.edu\\ }
}

\begin{document}
\bstctlcite{IEEEexample:BSTcontrol}

\maketitle
\conf{2021 IEEE High Performance Extreme Computing Conference (HPEC)}
\copyrightnotice
\begin{abstract}
    Iterative graph algorithms often compute intermediate
    values and update them as computation progresses.
    Updated output values are used as inputs for computations in current or subsequent iterations;
    hence the number of iterations required for values to converge can potentially reduce
    if the newest values are asynchronously made 
    available to other updates computed in the same iteration.
    In a multi-threaded shared memory system, the immediate propagation of updated values can cause
    memory contention that may offset the benefit of propagating updates sooner.
    In some cases, the benefit of a smaller number of iterations may be 
    diminished by each iteration taking longer.
    Our key idea is to combine the low memory contention that synchronous approaches have
    with the faster information sharing of asynchronous approaches.
    Our hybrid approach buffers updates from threads locally before committing 
    them to the global store to control how often threads may cause conflicts for others
    while still sharing data within one iteration and hence speeding convergence. 
    On a 112-thread CPU system, our hybrid approach attains up to 
    4.5\% - 19.4\% speedup over an asynchronous approach for Pagerank
    and up to 1.9\% - 17\% speedup over asynchronous Bellman Ford SSSP.
    Further, our hybrid approach attains 2.56x better performance than the synchronous approach.
    Finally, we provide insights as to why delaying updates is not helpful on certain graphs
    where connectivity is clustered on the main diagonal of the adjacency matrix.
\end{abstract}

\section{Introduction}
Many graph operations perform updates on the 
same vertices repeatedly.
In such algorithms a vertex's value is set, 
used as inputs in future computations, and later updated to a
more `correct' value.
Through multiple iterations, or rounds, the vertex 
values converge to some stopping point and 
the algorithm terminates. In this paper, we refer to these as \emph{iterative graph algorithms}.

Excluding algorithms that require some strict ordering on work
performed such as parent-list Breath First Search (BFS) 
and Dijkstra's SSSP~\cite{10.1007/BF01386390}, iterative graph 
algorithms may be implemented with
either synchronous or asynchronous approaches.
Operating synchronously, information generated during one 
iteration is made available
only at the start of the next iteration.
The advantage of such an approach is that it simplifies multi-threaded implementations. As long as the output vertex values are computed by different threads, each thread can perform the work assigned to it in a particular iteration without requiring synchronization with other threads. Only a single synchronization between all threads at the end of each iteration is required.
The drawback of such an implementation is that the vertex values in a single iteration are all restricted to the ``stale'' values obtained in the last iteration. This slows down the propagation of newly generated vertex values, as the new values are only available at the start of the next iteration.

In contrast, asynchronous implementations make information 
generated during a round immediately available to subsequent computations in the same iteration.
Data is propagated as fast as the 
underlying computer architecture and the graph topology allow. Hence, the algorithm may converge to a stop in fewer total iterations. 
%
%
A comparative work of multiple graph algorithms implemented across multiple graph libraries in part showed that asynchronous versions 
of Page Rank tended to out-perform synchronous versions~\cite{azad_evaluation_2020}.
Past and recent works also corroborate this observation~\cite{arasu_pagerank_2002, silvestre_pagerank_2018}.

However, the very property of sharing information immediately in asynchronous algorithms may correspondingly increase the time to compute each iteration in a multi-threaded implementation. For multi-threaded implementations on shared memory machines, newly computed data will be stored to a memory location that can be accessed by all threads. This may invalidate previously loaded cache lines, forcing threads reading data from those cache lines to reload. As such, while the number of iterations tends to be lower, the average time per iteration may increase.


\begin{table}[!tbp]
\caption{Number of rounds and average round time for Page Rank on GAP benchmark graphs, 32-thread Intel Haswell.}
\centering
\begin{tabular}{*7c}
\toprule
& \multicolumn{3}{c}{Rounds} & \multicolumn{3}{c}{Avg. Time per Round (s)}\\
\midrule
Graph & Synch & Asynch & Hybrid & Synch & Asynch & Hybrid\\
\midrule
Kron    & 7  & 5  & 5  & 2.94 & \textbf{3.02} & 2.92 \\
Road    & 39 & 18 & 20 & 0.02 & 0.02 & 0.02 \\
Twitter & 22 & 18 & 18 & 0.87 & 0.86 & 0.78 \\
Urand   & 6  & 4  & 4  & 4.06 & \textbf{4.11} & 4.02 \\
Web     & 29 & 19 & 22 & 0.16 & 0.15 & 0.14 \\
\bottomrule
\end{tabular}
\label{tab:pr_rounds_times}
\vspace{-5mm}
\end{table}

In this paper, we discuss the implementation of \emph{delayed asynchronous iterative graph algorithms}, a hybrid solution that combines the benefits of both synchronous and asynchronous algorithms. 
Specifically, our proposed delayed asynchronous algorithm reduces the number of individual writes in which updated information is propagated to all threads. This in turn reduces the potential memory conflicts that may result in increased execution time per round. A tunable parameter $\delta$ is introduced to  determine how often newly computed data is propagated. Setting $\delta$ to 0 makes our hybrid solution similar to a fully asynchronous algorithm, while setting it to ``infinity'' makes it a synchronous implementation.

We illustrate the benefits of the delayed asynchronous algorithms over both synchronous and asynchronous algorithms in Table~\ref{tab:pr_rounds_times} by  comparing the number of rounds and average time per round of the 3 styles of algorithms for computing Page Rank on the 5 graphs defined in the GAP benchmark~\cite{beamer_gap_2017}.
The asynchronous approach finishes in 2 to 21 fewer iterations over its synchronous counterpart. However, in the case of Urand and Kron, the average time to compute each round increases (bolded). For the other graphs, the time per round remains about the same. Our proposed hybrid approach similarly reduces the number of rounds as the asynchronous approach. In addition, each round of our hybrid solution takes the least time among the three solutions.
This yields better overall runtimes than the other two approaches for 3 of 5 graphs.



{\bf Contributions: } 
The contributions of this paper are:
\begin{itemize}
\item We introduce \emph{delayed asynchronous iterative graph algorithms}, a hybrid that combines the benefits of both synchronous and asynchronous graph algorithms.

\item We discuss the factors that affect the choice of different values of $\delta$, the delay parameter, which allows one to design an implementation that is a hybrid between 
the two extremes, i.e. 
synchronous and asynchronous algorithms.




\item We demonstrate the performance benefits of the delayed asynchronous algorithm where 
the performance improvement of Page Rank on moderately-threaded CPU 
(32 threads) is between 2.4 and 9.5\%, and between 4.5 - 19.4\% on a
highly parallel CPU (112 threads).  
On Bellman-Ford Single Source Shortest Paths, the improvement is between 1.9 and 17\% for 112 threads.
\end{itemize}

\section{Background and Motivation}
\subsection{Synchrony and Asynchrony in Graph Algorithms}
Synchronous and asynchronous graph algorithms are discussed extensively in graph literature. Specifically for Page Rank, a common approach is to contrast 
Gauss-Seidel iterative computation
with Jacobi synchronous iterative computation~\cite{chazan_chaotic_1969, silvestre_pagerank_2018, arasu_pagerank_2002} based on linear algebraic methods.

Architecturally, many works on asynchronous graph algorithms focus on distributed execution, where iterative, bulk-synchronous parallel execution is considered~\cite{valiant_bridging_1990}.
Some works study completely asynchronous execution policies 
without distinct compute iterations, or super-steps~\cite{firoz_runtime_2018, firoz_synchronization-avoiding_2018}.
These works focus on removing points of synchronization and focus
entirely on asynchronous execution in a distributed environment.

Xie et al. combine super-step-free asynchrony with synchronized iterative 
operation, but do not algorithmically combine the two; instead they switch 
between two distinct modes using a heuristic~\cite{xie_sync_2015}. 

Similar asynchronous execution is explored on single-node shared memory systems using thread-local work queues~\cite{pearce_multithreaded_2010} and bulk-asynchronous parallel execution on heterogeneous systems~\cite{dathathri_gluon-async_2019}.
These works focus on asynchronous execution, without considering the benefits of hybrid synchronous-asynchronous behavior.

Vora et al. address asynchronous execution in a shared-memory distributed setting, wherein high inter-node latency may cause excessive use of stale data without best-effort refresh policies~\cite{vora_aspire_2014}. 
By contrast, we reduce cache line invalidations in a single-node shared memory system by delaying writes to shared memory and thus increasing use of stale values in a controlled manner.

Overall, our work differs from those described in that we seek a more general approach for synchronous and asynchronous execution to apply to Page Rank and other graph operations. 
Our approach specifically combines both execution styles within the same iteration 
to enable flexibility in blending their benefits.
We apply this algorithmic innovation to iterative graph algorithms on single-node shared-memory systems, 
where the previous works have focused on either distributed systems, non-hybrid asynchronous execution, or both.

\begin{figure*}[ht!]
    \centering
    \includegraphics[width=0.8\textwidth]{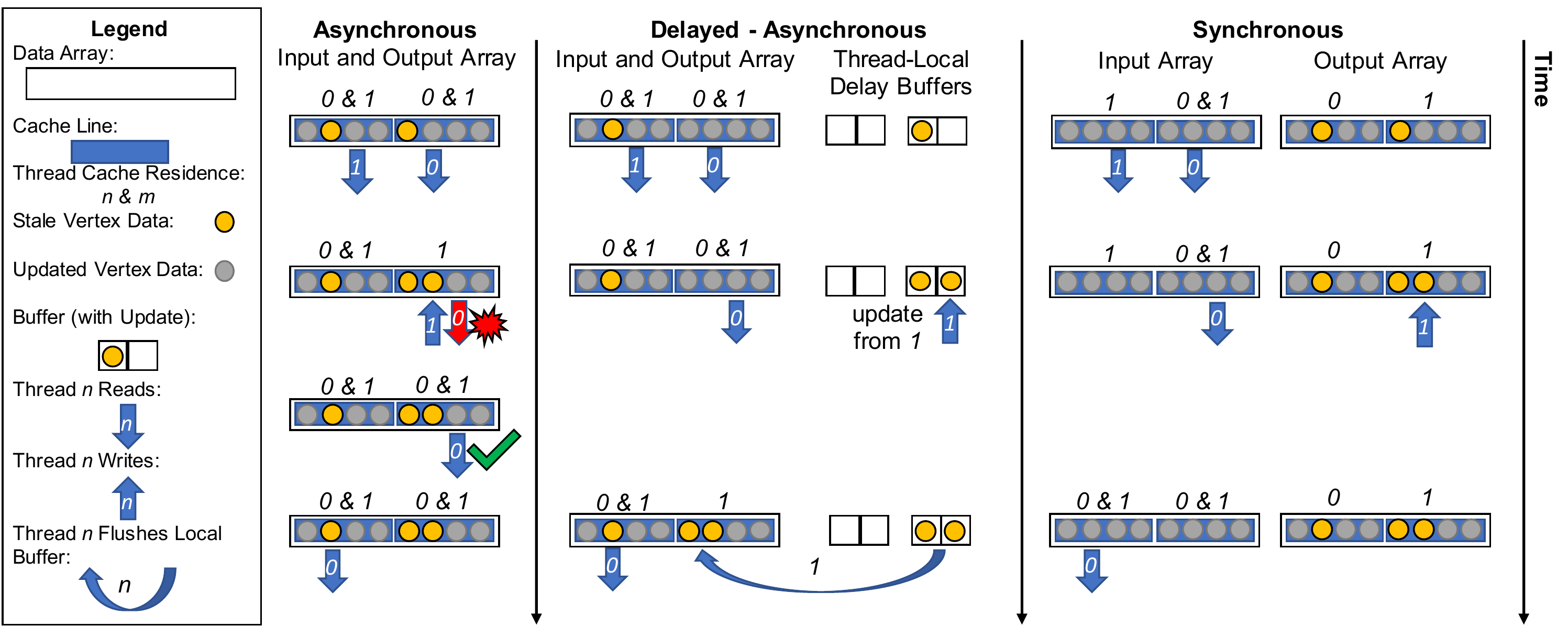}
    \vspace{-3mm}
    \caption{Legend and three panels showing two threads reading and writing two cache lines of data in asynchronous (left), hybrid delayed asynchronous (middle), and synchronous (right) modes. Yellow dots represents newly computed values; grey dots are old values. Asynchronous execution incurs a cache line invalidation for thread 0 when thread 1 writes into a cache line previously read by thread 0. The other two implementations avoid the invalidation by writing to an external buffer. Delayed asynchronous makes the updated values available faster than synchronous execution. 
    }
    \label{fig:graph_algo_components}
    \vspace{-3mm}
\end{figure*}

\subsection{Hybrid Graph Algorithms}
A small group of graph algorithms are designed through the blending of two or more ``classical'' algorithms. Two examples of such hybrids are the direction-optimizing breath first search (DO-BFS)~\cite{beamer_direction-optimizing_2012} and delta-stepping single source shortest paths (DS-SSSP)~\cite{meyer_delta-stepping_2003}. 

DO-BFS is derived from the classical top-down and bottom-up approaches, sometimes referred to as push and pull-style~\cite{beamer_direction-optimizing_2012, yang_implementing_2018}. At the end of each iteration of DO-BFS, a heuristic decides if the push or pull implementation will be used in the next iteration. In this case, the algorithm uses both classical algorithms separately such that only one is invoked during each iteration.

DS-SSSP is a single algorithm that is a hybrid between two classical SSSP algorithms, Dijsktra SSSP~\cite{dijkstra} and Bellman-Ford~\cite{bellman_routing_1958}. Unlike DO-BFS, DS-SSSP blends the characteristics of the two classical algorithms through setting of the parameter $\Delta$.
Setting the parameter $\Delta$ to zero, DS-SSSP is functionally equivalent to Dijkstra SSSP. When $\Delta$ is set to infinity, DS-SSSP is equivalent to the Bellman-Ford algorithm. Typically, the step-size $\Delta$ is graph-dependent. Recently, it has been shown that allowing the value of $\Delta$ to change between different iterations can attain better performance~\cite{blelloch_parallel_2016}.

The approach taken by DO-BFS discretely switches between existing algorithms.
Chakravarthy et al. use this same approach in combining direction optimization and mode switching between DS-SSSP and Bellman-Ford to speed convergence~\cite{chakaravarthy_scalable_2014}.


In contrast, our delayed asynchronous algorithm is similar to the DS-SSSP algorithm in that the parameter, $\delta$, controls how much of the characteristics of either synchronous or asynchronous algorithms is present in the final implementation.
The parameter $\delta$ allows us to fine-tune the hybrid nature of our approach without a discrete mode-switch and yields a novel hybrid algorithm.

\section{Delayed Asynchronous  Graph Algorithms}
Our proposed delayed asynchronous algorithm is similar to the synchronous algorithm in that computed values are first written into output buffers separate from the input buffer that is shared globally. When these output buffers are full, they are flushed and data is written into the global input buffer to be used by all threads. 
By setting the output buffer size to a value that is smaller than the number of vertices assigned to a thread, the delayed asynchronous algorithm can be made to perform flushing of the data more than once per iteration. The flushing of the output buffer multiple times makes newly computed values available before the end of the current iteration; making the proposed algorithm a hybrid of both the synchronous and asynchronous algorithms.

We illustrate how our delayed asynchronous algorithm is a hybrid of both the synchronous and asynchronous algorithms in Figure~\ref{fig:graph_algo_components}.
Vertex data mapping into two adjacent cache lines is read and written by 
two threads in an asynchronous fashion.
In the presence of one writer to the second cache line, thread 0 is unable to access any data stored in that line.
Only after thread 1 finishes writing can thread 0
access data in the same cache line.
Hence, even having a single writer among readers in 
asynchronous execution can cause delays for threads 
reading data in order to compute their own vertex updates.

In the center of Figure~\ref{fig:graph_algo_components}, the write 
that caused a cache line invalidation for thread 0 is delayed by thread 1, allowing thread 0 to proceed using stale data.
Delaying updates in this way can reduce write-read contention with other threads,
in the same way that fully synchronous execution avoids such conflict by writing 
all new values to a completely separate array from the one which threads
read from (right side of the figure).
However, in the hybrid approach, updates are still made available to other threads at some point during an iteration,
as in the center panel where thread 0's last access is to data updated at some point earlier in the same iteration.
\subsection{Design decisions}
Our hybrid solution is designed to reduce memory overheads arising from 
cache line invalidation due to the coherence protocol in multi-threaded asynchronous execution. 
On modern shared memory systems, cache lines are invalidated when a thread updates a value in a cache line that has been read by multiple threads. 
To reduce the number of cache line invalidations, the following decisions were made for our implementation:
\begin{itemize}
    \item {\emph{Pull-style implementations.}} 
A pull-style algorithm, where each vertex value is updated by one and only one thread, is utilized. This prevents the situation where multiple threads update the same memory location and cause excessive cache invalidation due to multiple writes.
    \item {\emph{Separate thread-local output buffer.}} Similar to synchronous algorithms, the computed outputs are written to a separate buffer from the inputs. As inputs are in a separate memory location, writes to the output will not invalidate data that was previously read since no cache lines are shared.  For our implementation, the output buffers are separate and thread-local.
     \item {\emph{Blocked partitioning of work.}} The work to be computed is partitioned amongst all threads in a contiguous blocked fashion using the given vertex IDs. Vertices are allocated to individual threads in a way that balances the aggregate number of in-neighbors per thread as much as possible. This allows us to allocate contiguous chunks of memory for storing the outputs of each thread. Furthermore, this blocked partitioning also reduces the number of cache lines that are ``dirtied'' when  the updated vertex values are pushed to the global memory space. For ease of implementation, our implementation uses a static partitioning of the vertices across all iterations of the algorithm.
\end{itemize}

\subsection{Delaying propagation with parameter $\delta$}


The delayed asynchronous approach introduces a parameter $\delta$ to control how much delay is applied to threads' updates.
Based on $\delta$, updates are stored locally to the thread and only copied into the global memory location when either 1) the buffer capacity determined by $\delta$ is exhausted, or 2) the thread has completed all assigned work for that iteration.

When  $\delta$ is set to 0, there is no delay buffer, and data is immediately written out to the global buffer. This is equivalent to the asynchronous algorithm. Similarly, when $\delta$ is set to the number of output values assigned to each thread, then all threads can store all their outputs before having to flush the buffer. This is equivalent to the synchronous version.

$\delta$ is sized in vertex data elements to a multiple of the cache line size so that 
flushing a full buffer makes maximal use of
bringing a cache line in from a further level of cache.
Finally, coalesced updates provided by an aligned buffer enable use of aligned vector loads. We discuss other factors affecting the choice of $\delta$ in Section~\ref{sec:factors}.

\subsection{Local vs Global Reads}
While we have focused on reducing memory conflicts when storing newly computed values, we highlight that there are two different variants for reading vertex values.

A simple implementation of the hybrid implementation is for threads to read from the global memory space. Newly updated values will not be available until individual threads flush them into the global memory space.
As asynchronous algorithms speed up convergence by propagating newly computed values as quickly as possible, an alternative implementation is to allow individual threads to read values previously computed but not flushed from their own delay buffer. This allows even faster information propagation. As the delay buffers are thread-local, there will not be memory conflicts.


In testing this alternate implementation, it is rarely faster than reading from the global memory space. For the rest of this paper, we only discuss results for global reading. 

\section{Factors affecting delay parameter $\delta$ }
\label{sec:factors}
Picking good values for $\delta$, the delay parameter that blends
the implementation between fully asynchronous and fully synchronous, 
depends on the platform, input, and algorithm. 
We examine the different conditions that affects how the parameter $\delta$ should be selected. In this section, we examine the following factors:
\begin{enumerate}
    \item Number of threads
    \item Connectivity of graphs (Graph Topology)
    \item Average number of updates in an iteration.
\end{enumerate}


We evaluate pull-based Page Rank and Bellman Ford SSSP.
For each algorithm, we test using the GAP benchmark graphs~\cite{beamer_gap_2017}.
Details of the graphs are shown in Table~\ref{tab:gap_graphs}. We report the performance of synchronous, asynchronous, and multiple
$\delta$ set-points for delayed asynchronous to determine which has 
the best performance for otherwise identical run-time conditions 
(same work allocation, unconditionally storing updates).
We test values of $\delta$ in power-of-two sizes from 16 up to 32768 elements, where each element is 32-bits.
Page rank scores are computed with floating point values and SSSP path lengths are computed with 32-bit unsigned integers.

For both workloads, each instance (synchronous, asynchronous, delayed asynchronous) is run 
for three trials and the average is taken.
In Page Rank, each trial runs to a convergence criterion such that the total absolute page rank 
score change across vertices from the penultimate iteration totals 1e-4.
For SSSP, the stopping criteria is that no update was generated in the last iteration.

\begin{table}
    \centering
    \caption{Statistics of GAP Benchmark Graphs}   
    \begin{tabular}{cccc}
    \toprule
    \bf Graphs     &  \bf Vertices &\bf Edges & \bf Symmetric? \\
    \midrule
    Road	&   57.7 M    &	 23.9 M & yes\\
    Urand	&   4,295.0 M &	134.2 M & yes\\
    Kron    &	4,223.3 M &	134.2 M & yes\\
    Twitter	&   1,468.4 M &	 61.6 M & no\\
    Web	    &   1,930.3 M &	 50.6 M & no\\
    \bottomrule
    \end{tabular}
    \label{tab:gap_graphs}
    \vspace{-5mm}
\end{table}

All experiments were performed on two Intel architectures: dual-socket Haswell (Xeon E5-2667 v3)
with 16 cores and 32 threads at 3.2 GHz and 
dual-socket Cascade Lake (Xeon Platinum 8280) with 56 cores and 112 threads at 2.7 GHz.

\subsection{Delaying Dense Updates in Many-Threaded Page Rank}
At each iteration of Page Rank, a vertex value is likely to change
because it is computed as a weighted combination of scores among incoming neighbors.
A change in even one in-neighbor's score will change the vertex score
and hence timeliness of any new information matters.
For such dense updates and at the largest number of threads on each platform, 
fully asynchronous execution may incur a large number of conflicts as each thread 
writes to the vertices assigned to it, potentially invalidating
a cache line for other threads repeatedly as it generates updates for 
each vertex score in the cache line one at a time.
Therefore, the benefit of the hybrid approach should be clearly visible at the 
maximum number of threads, because it balances the benefit of sharing updates 
earlier with a reduction in number of write-accesses to shared cache lines.

\begin{figure}
    \centering
    \includegraphics[width=\linewidth]{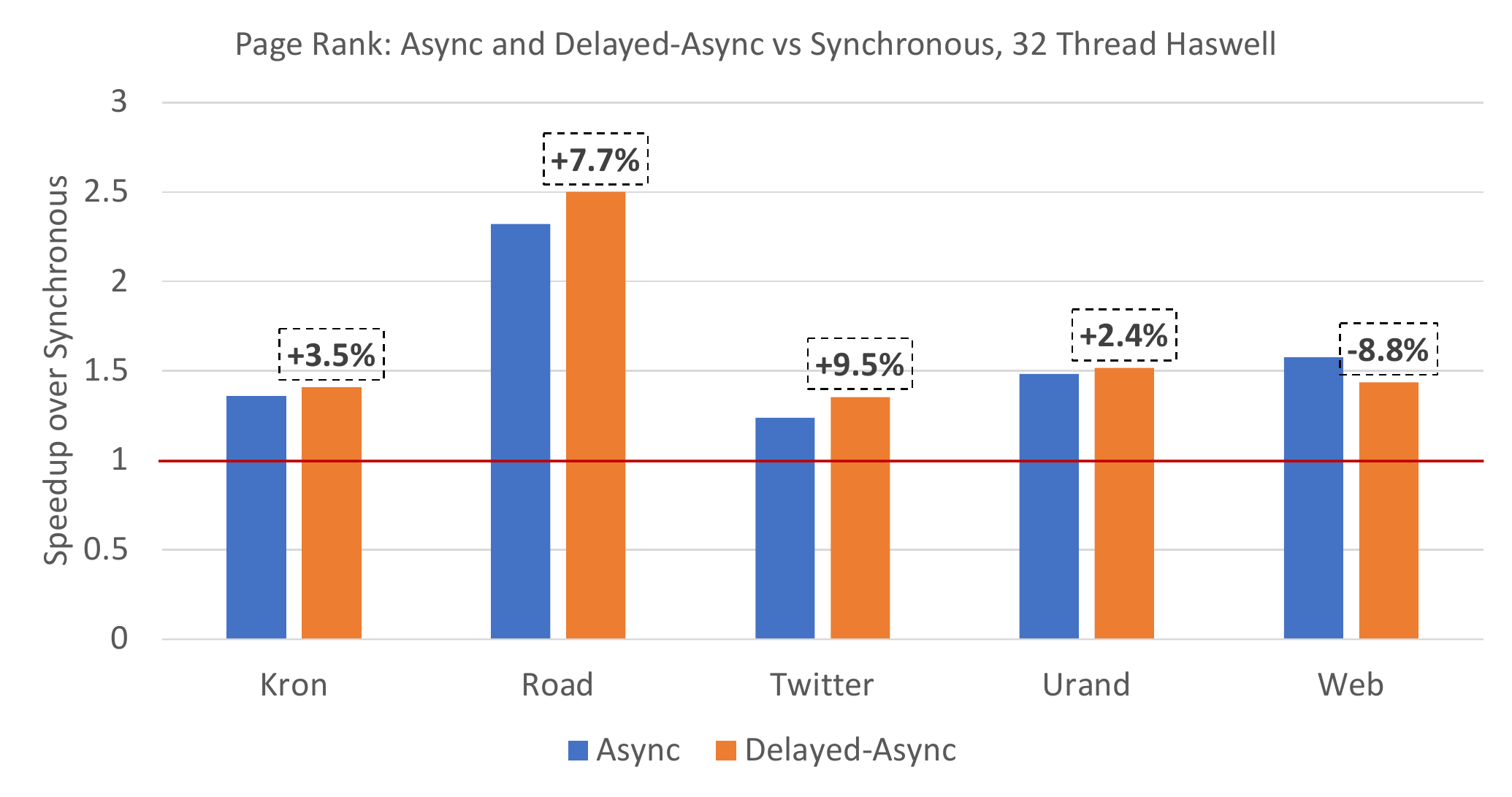}
    \includegraphics[width=\linewidth]{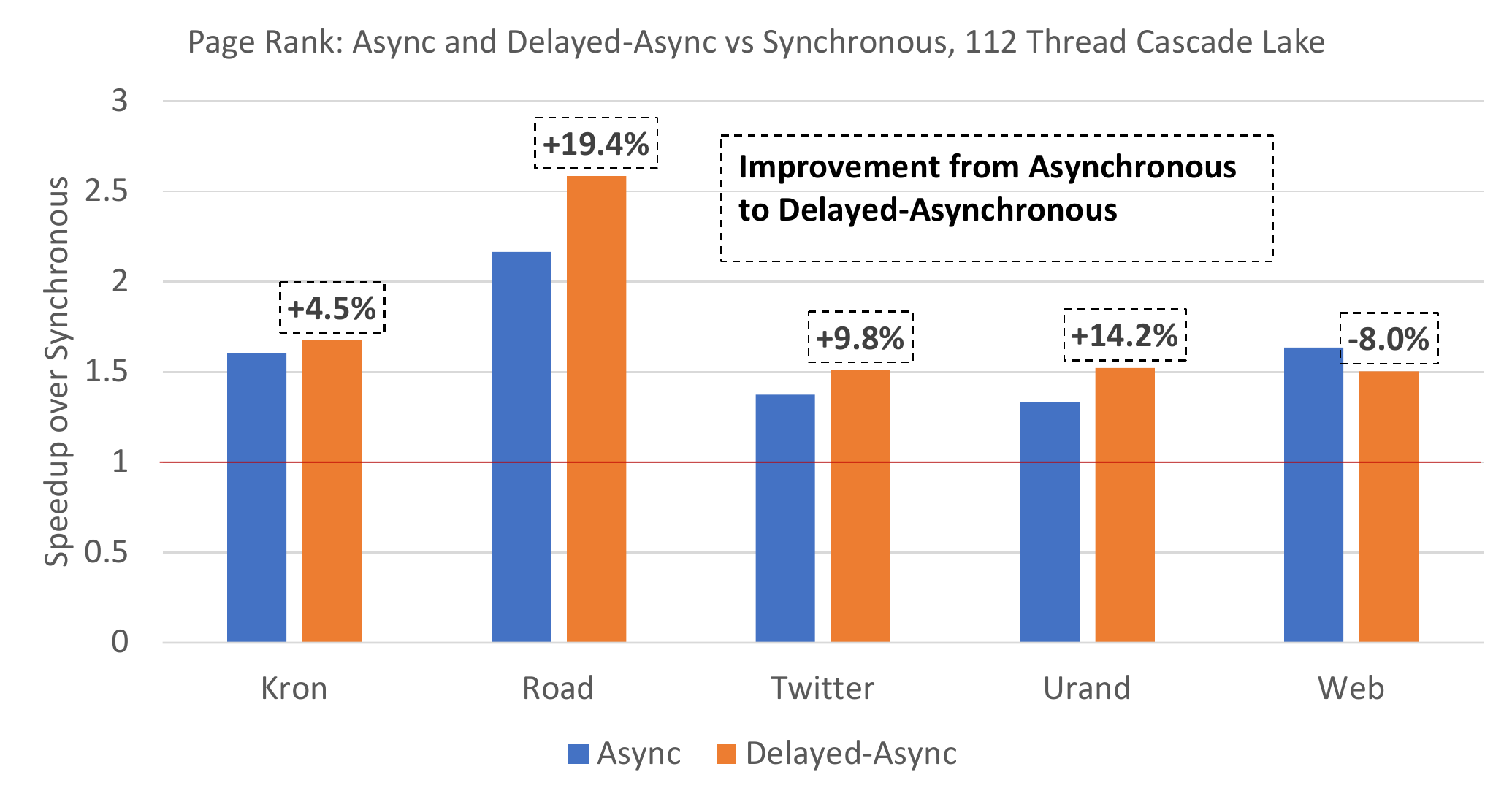}
    \vspace{-9mm}
    \caption{Page Rank speedup over synchronous baseline (red line), for asynchronous and delayed asynchronous implementations. For each graph except web, delaying updates improves performance over the asynchronous approach, which in turn is always better than synchronous execution. 
    }
    \vspace{-6mm}
    \label{fig:pr_results}
\end{figure}

Figure~\ref{fig:pr_results} confirms these two points concerning 
the performance of asynchronous and delayed asynchronous implementations, 
normalized against synchronous performance. 
First, asynchronous and hybrid execution always offer better performance 
compared to the baseline synchronous implementation.
This confirms that sharing information sooner, especially in the 
dense update environment, leads to faster convergence.
Second, delaying updates often offers a performance improvement over the asynchronous approach
by consolidating writes from threads.
The improvement over asynchronous execution across all GAP graphs is between 
2.4\% to 9.5\% on the Haswell system and between 4.5\% to 19.4\% on the 
Cascade Lake system with all threads active.


\subsection{Scaling Threads with Dense Updates in Page Rank}
As the number of threads increases for these graphs, the amount of work per thread decreases
and inter-thread communication may increase. 
Hence we would expect to see larger values of $\delta$ be effective for 
a low number of threads, while a smaller value of $\delta$ would be better at a high 
thread count because it shares information faster.

For the largest thread setting in the scaling study, we use all available hyper-threads on the machine. 
Both machines are dual-socket, so the next lowest thread setting - half of the thread complement - 
is arranged across sockets to avoid hyper-threading.
Below this, the thread count is progressively halved and they are pinned with their memory on one socket.

\begin{figure}[t]
    \centering
    \includegraphics[width=\linewidth]{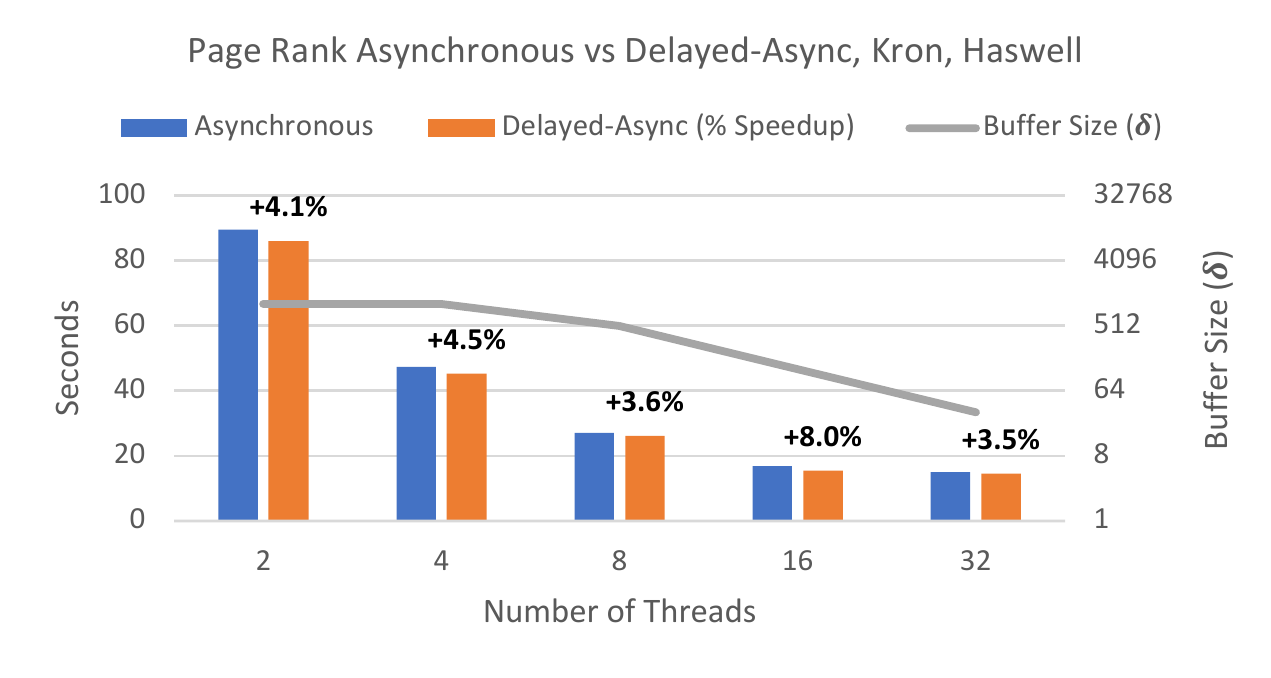}
    \includegraphics[width=\linewidth]{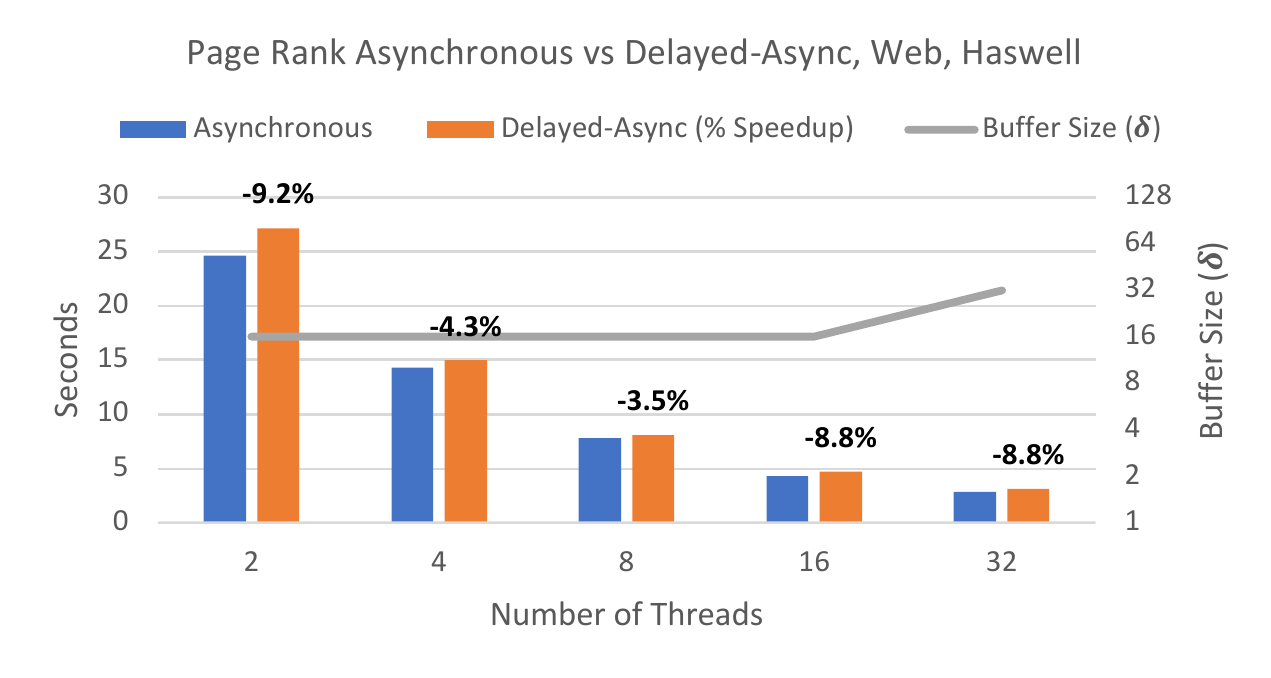}
    \vspace{-10mm}
    \caption{
    Thread Scaling for Pagerank on up to 32 threads, Haswell system.
    For Kron the best buffer sizes (value of $\delta$) trend downward as thread count increases. 
    On Web, even the smallest tested $\delta$ value (16) does not achieve improvement over asynchronous execution.
    }
    \label{fig:pr_scaling_hsw}
    \vspace{-6mm}
\end{figure}

\begin{figure}[t]
    \centering
    \includegraphics[width=\linewidth]{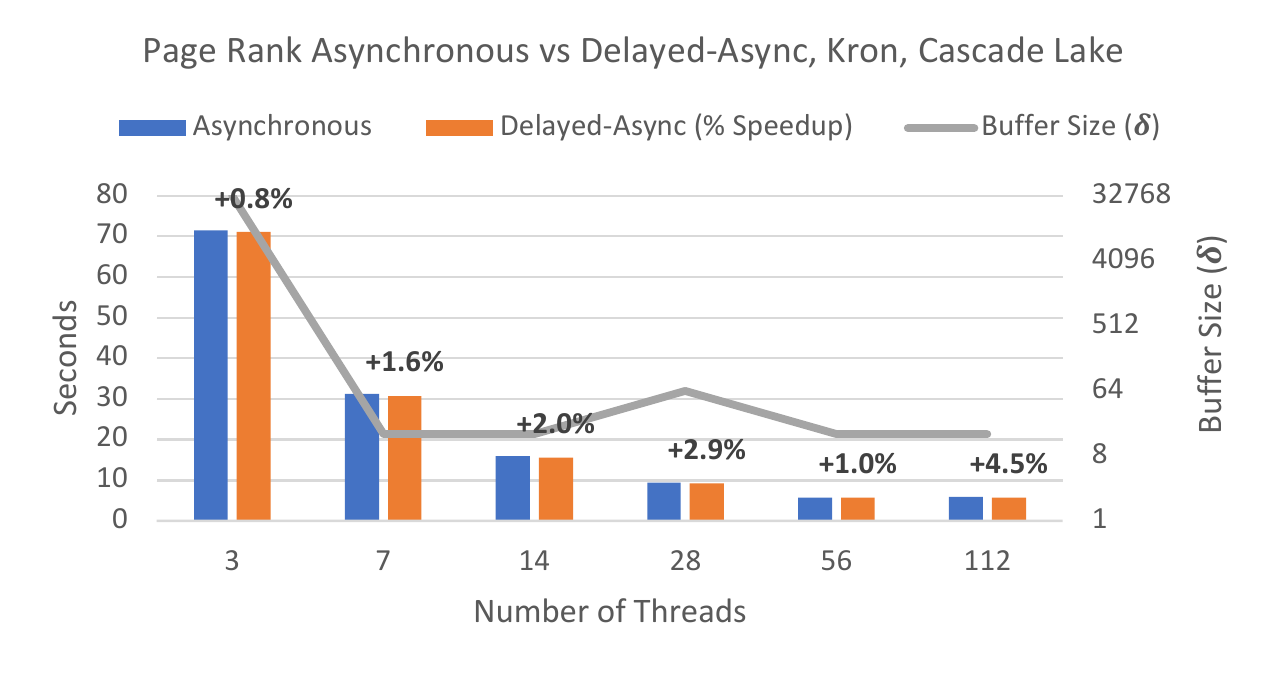}
    \includegraphics[width=\linewidth]{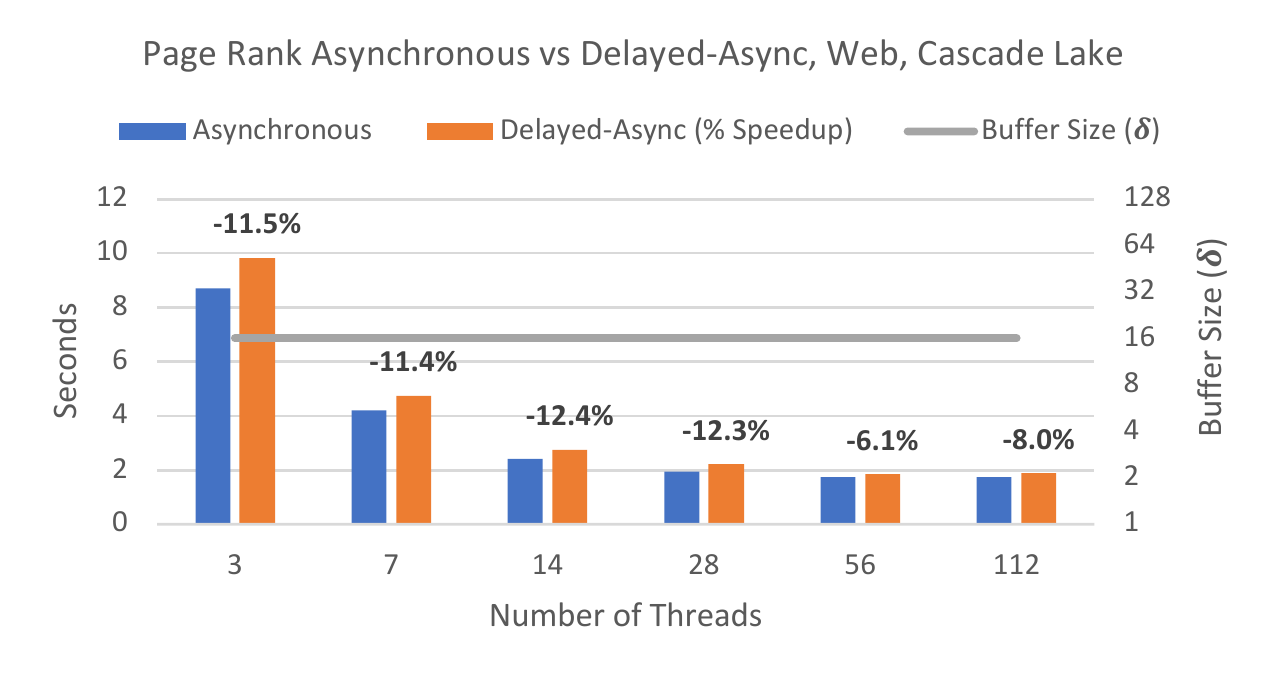}
    \vspace{-10mm}
    \caption{
    Thread Scaling for Pagerank on up to 112 threads, Cascade Lake system. 
    Kron 
    shows improvement from delaying updates with decreasing $\delta$ as thread count rises.
    On Web there is no benefit and the penalty is smallest for a relatively small $\delta$ across all thread counts.
    }
    \label{fig:pr_scaling_clx}
        \vspace{-6mm}
\end{figure}

Figures~\ref{fig:pr_scaling_hsw} and~\ref{fig:pr_scaling_clx} show the runtime of the delayed and asynchronous approaches
at each thread configuration for Kron and Web.
Detailed plots for Twitter, Urand, and Road are omitted due to space constraints.
The value of $\delta$ that achieves the annotated stated speedup over asynchronous is shown with a grey line on the secondary axis.

Indeed, on Kron, the best $\delta$ decreases as the number of threads increases, for two likely reasons.
First, less work per thread for a large number of threads implies fewer updates generated per thread.
Second, updates must be communicated sooner to avoid slowing convergence in the presence of many threads
and the long range connections present in Kron. 
This is consistent with requiring more information transfer, while
still needing some delay in updates to avoid excessive conflict between threads reading and writing.

%

For Web, the best value of $\delta$ is often the smallest one tested (16 elements),
and delayed asynchronous execution is always worse than asynchronous.
As we discuss in the next subsection, this is attributable to the Web's topology.

\label{subsec:graph_topo}
\begin{figure}[th!]
    \centering
    \includegraphics[width=0.50\linewidth]{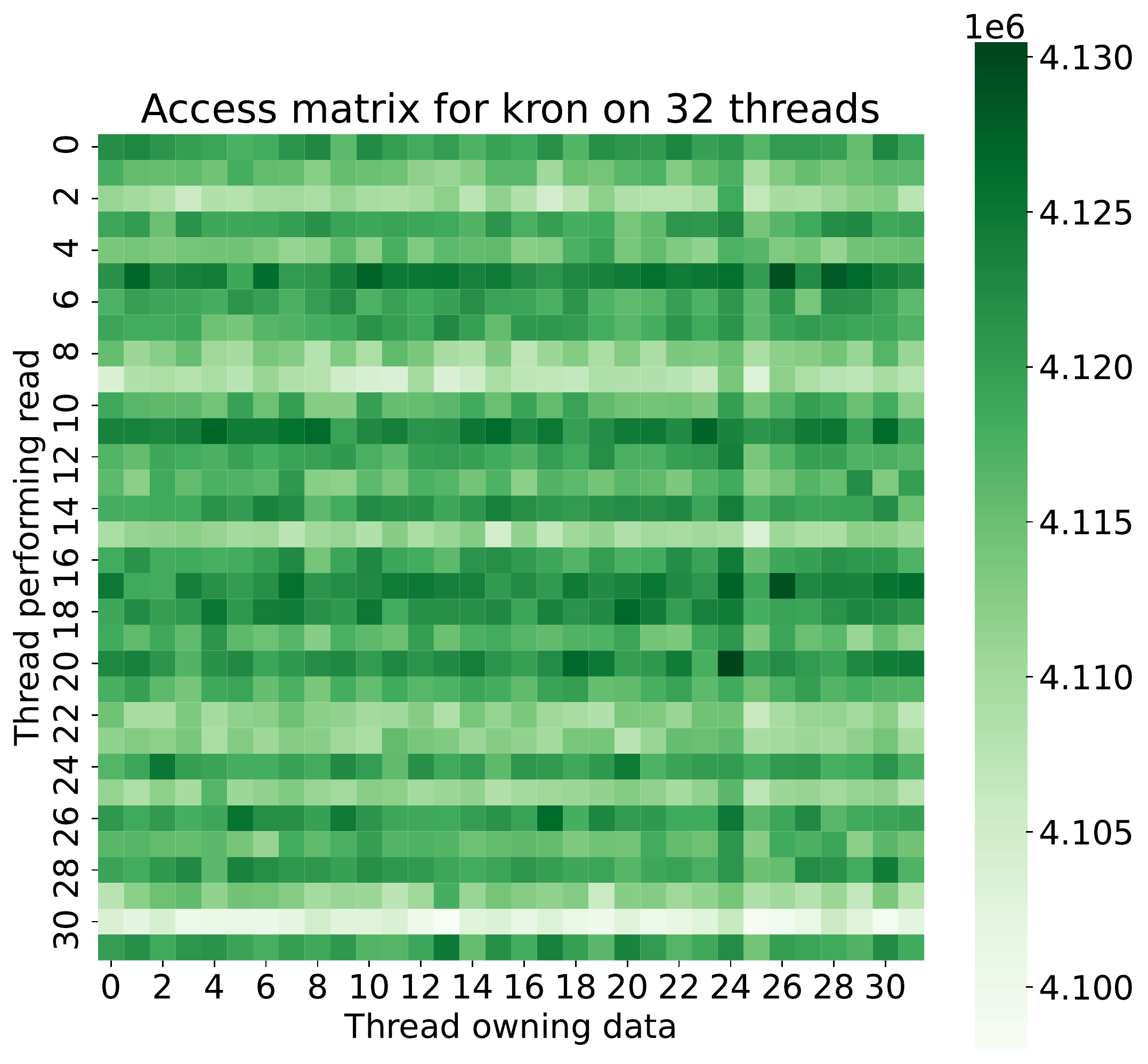}
    \includegraphics[width=0.47\linewidth]{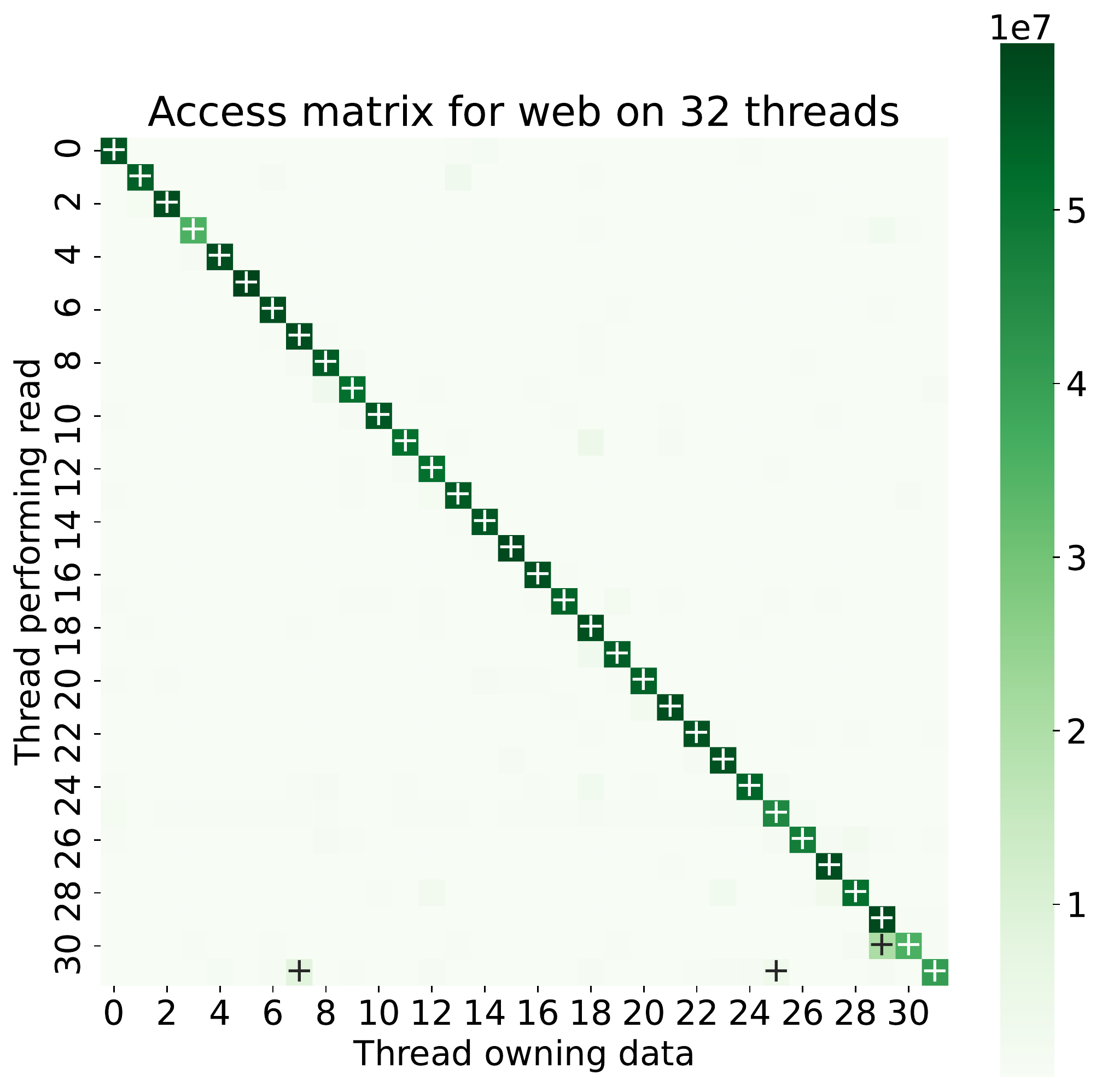}
    \caption{The access matrices show the proportion of memory accesses made from each thread (row) to other threads (column) in a 32-thread setup (Haswell). A plus in the box indicates that the row thread received at least 6.25\% or (1/32) of accesses from itself rather than other threads. Kron does not show significant imbalance in local vs remote accesses. Web shows high clustering on the main diagonal. This indicates that a thread references its own updated information far more than it references information generated by other threads. 
    }
    \label{fig:access_matrices}
    \vspace{-5mm}
\end{figure}

\subsection{Effects of Graph Topology} 
Graph topology is expected to affect the amount of communication 
between threads, which in turn affects the potential for write-read 
contention on cache lines. 
Here we study thread communication induced by the topologies of  
Kron and Web.

Recall that in each of these experiments, we statically assign vertices
to threads based on their degree in contiguous blocks.
In one round of Page Rank, we instrument the number of reads by each thread into every other thread's assigned vertex data.
This effectively creates a coarsened adjacency matrix of the graph
with partitions based on the static work assignment.

Figure~\ref{fig:access_matrices} shows the number of accesses made 
to data that a thread will change itself (local) v.s. 
reads on data that another thread will write in that round.
The row index corresponds the thread ID reading, and the 
column index corresponds to the thread ID that `owns' (writes) that data.
In spite of the balanced number of edges assigned to each thread, 
some graph topologies cause a larger proportion of local compared to remote accesses.

Kron and Web are both scale free.
However, while Kron exhibits relatively diffuse reads from all threads,
Web exhibits dense clustering in which the thread writing
some range of vertex data is a major reader of that data. 
Figure~\ref{fig:access_matrices} shows that the darkest 
read-write regions are on the main diagonal.
As updates are used mainly by the thread that creates them,
there is little benefit in delaying the global write-out because doing so does not relieve 
inter-thread memory contention.

\subsection{Effects of average number of updates in each iteration}
\begin{figure}[t!]
    \centering
    \includegraphics[width=\linewidth]{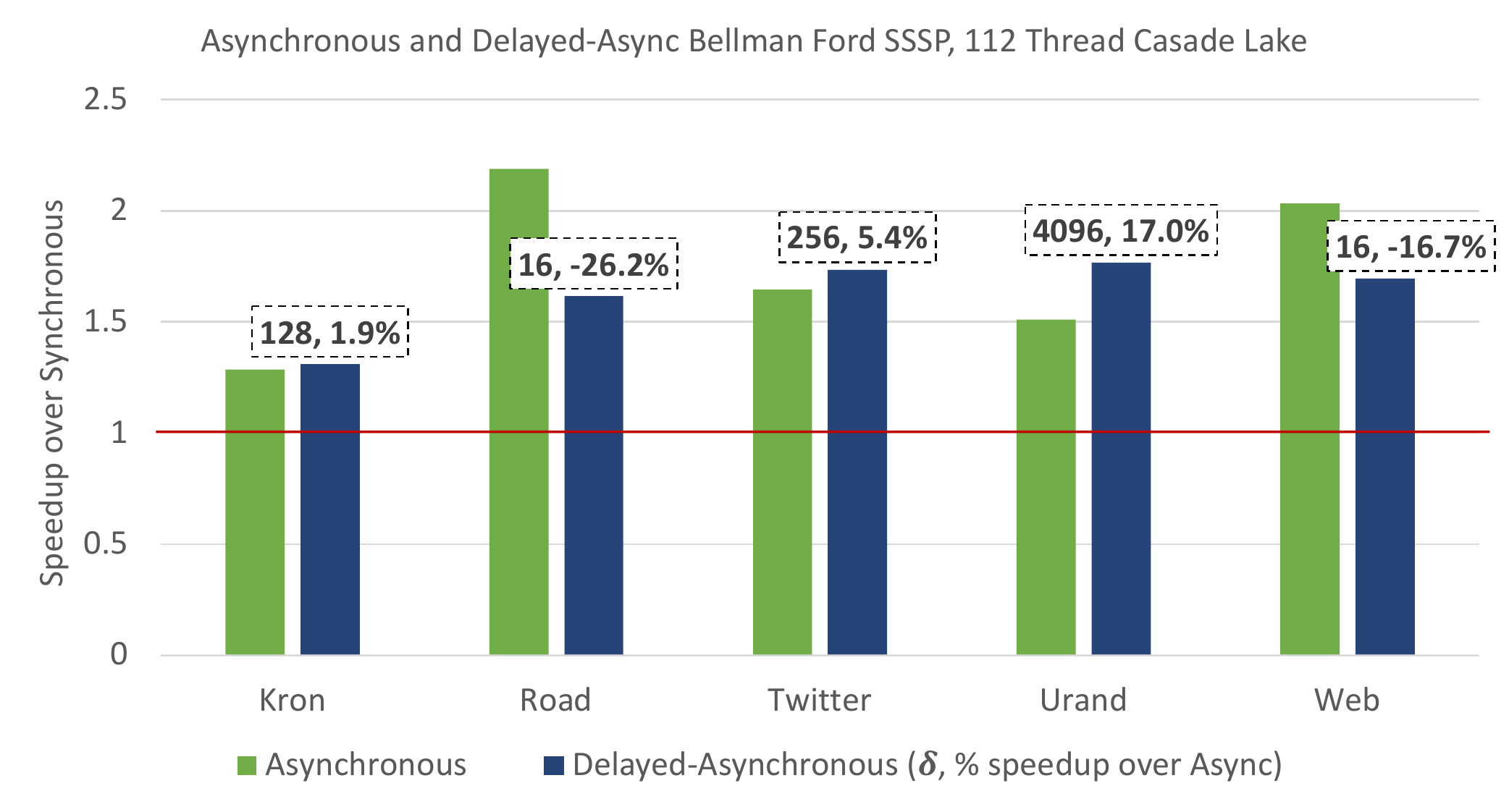}
    \vspace{-5mm}
    \caption{Speedup of asynchronous and delayed asynchronous SSSP implementations over baseline synchronous (red line) on 112-thread Cascade Lake. Kron, Urand, and Twitter benefit from they hybrid approach, while Road and Web graphs do not. Unlike Page Rank, in SSSP, fewer updates are made on average per round.}
    \label{fig:sssp_results}
    \vspace{-5mm}
\end{figure}
We demonstrate that a lower average number of updates in each iteration decreases the benefits of our hybrid approach.

Bellman Ford results were collected using the given weights for each of the GAP graphs.
In this analysis, we expect to see that a smaller $\delta$ is better for SSSP because fewer updates are needed as the algorithm progresses and thus every update created is of greater significance to convergence.

Indeed, Figure~\ref{fig:sssp_results} shows that delaying updates is effective 
on SSSP for only some of the graphs in the GAP collection, 
while it was effective on Page Rank for most of them.
Web and Road demonstrate poor performance when buffering is used for SSSP.
For Web, this is likely due to the very localized access pattern 
shown in Figure~\ref{fig:access_matrices}.
For Road, delaying updates is not beneficial because doing so slows down 
information transfer in a graph that already has slow information 
transfer due to large diameter and very low average degree.
Kron, Urand, and Twitter do show some benefit from the hybrid approach,
and these graphs are the ones with more long-range connections and 
without the dense clustering in Web.


\section{Conclusion}
We propose the delayed asynchronous approach for iterative graph algorithms. Similar to hybrid algorithms such as delta-stepping SSSP, our implementation spans the space of possible implementations between synchronous and asynchronous behavior through a delay parameter $\delta$. 

We discuss factors affecting the choice of $\delta$ and demonstrate the effects of these factors through a detailed evaluation using PageRank and Bellman-ford SSSP. We show that on highly threaded systems, the performance of the delayed asynchronous implementation can be up to 19.4\% faster than an asynchronous implementation, and up to 2.56x faster than a synchronous implementation.


Our evaluation shows that delay-buffered Page Rank achieves reasonable 
performance improvements (2.4 - 19.4\%) over asynchronous execution
and that these results are best at high thread counts.
Graphs with topologies not amenable to buffering are found to 
have high levels of local clustering.
This analysis of a graph's topology can be precomputed, 
giving a potential way to determine when to buffer in practice.

For Bellman Ford SSSP, we show modest speedup from buffering 
between 1.9 and 17\% on the 112-thread machine when
the graph is amenable to buffering. These findings suggest that
the reduced number of updates in SSSP requires more care in when 
and on what graphs to buffer vertex information updates.

In future work, we would extend the idea of buffering to other 
pull-style algorithms, 
including where updates may only be conditionally written.
We would also consider other ways of assigning vertices to threads 
to balance the thread-local and thread-remote reads.
Finally, further work must be done to determine what buffer size to use,
dependent on both the graph's topology and the number of threads on the system.

\section*{Acknowledgement}
This material is based upon work funded and supported by the Department of Defense under Contract No. FA8702-15-D-0002 with Carnegie Mellon University for the operation of the Software Engineering Institute, a federally funded research and development center.
[DISTRIBUTION STATEMENT A] This material has been approved for public release and unlimited distribution.  Please see Copyright notice for non-US Government use and distribution.
DM21-0741

Mark Blanco is supported by the National Science Foundation Graduate Research Fellowship Program under Grant No. DGE 1745016, and the Jack and Mildred Bowers Endowed Fellowship and the Benjamin Garver Lamme and Westinghouse Graduate Fellowship.
Any opinions, findings, and conclusions or recommendations expressed in this material are those of the author(s) and do not necessarily reflect the views of the National Science Foundation.

Use of the Cascade Lake system was provided by Henry Gabb and Ramesh Peri at Intel, and we are grateful for their support.

\bibliographystyle{IEEEtran}
\bibliography{refs}

\end{document}